\documentclass[10pt]{article}
\usepackage{graphicx}

\begin{document}

\title{Dependence of far-field characteristics on the number of lasing
modes in stadium-shaped InGaAsP microlasers}

\author{Muhan Choi\thanks{Corresponding author: choi@atr.jp}, Susumu
  Shinohara, and Takahisa Harayama\\
Department of Nonlinear Science, ATR Wave Engineering Laboratories, \\
2-2-2 Hikaridai, Seika-cho, Soraku-gun, Kyoto 619-0288, Japan}
\date{}
\maketitle

\begin{abstract}
We study spectral and far-field characteristics of lasing emission
from stadium-shaped semiconductor (InGaAsP) microlasers. We
demonstrate that the correspondence between a lasing far-field
emission pattern and the result of a ray simulation becomes better
as the number of lasing modes increases. This phenomenon is
reproduced in the wave calculation of the cavity modes.
\end{abstract}


\section{Introduction}

Two-dimensional (2D) optical microcavities have been attracting the
attention of many researchers, on one hand because of their
potential applications in optical engineering \cite{Vahala03} and on
the other hand because they offer an experimental testing ground for
the quantum/wave chaos theory \cite{Nockel97}; the basic formalism
for the wave description of a quantum billiard can be applied with
some little modifications to the description of the light field
confined in a 2D optical microcavity \cite{Schwefel05}.
One of the simplest cavity shapes that are interesting from the
viewpoint of the quantum/wave chaos theory is the shape called the
Bunimovich stadium, as illustrated in the inset of Fig.
\ref{fig:spectrum}(a).
This shape is well known for its mathematically proven property that
a billiard ball or a ray inside the cavity exhibits fully chaotic
dynamics \cite{Bunimovich77}.

The idea of using the stadium shape as a laser cavity has been
examined both theoretically and experimentally \cite{SB,Shinohara06,
  Shinohara08, Fukushima04, Fang07, Park07, Lebental06}.
Actual stadium-shaped microcavities have been fabricated using
materials such as semiconductors \cite{Shinohara08, Fukushima04,
  Fang07, Park07} and polymers \cite{Lebental06}.
A remarkable property that is commonly observed for relatively large
cavity sizes is that the lasing emission patterns can be well
explained by a ray model \cite{Shinohara08, Fukushima04,
Lebental06}.
The ray model can be regarded as an open Hamiltonian system, whose
openness, that is the light leakage at the cavity boundary, is
described by Fresnel's law \cite{Nockel97, Schwefel05}.
A theoretical mechanism leading to the correspondence between ray
and wave descriptions has been numerically investigated in Refs.
\cite{Shinohara06, Shinohara08}.
In the present paper, we report experimental results on the
dependence of far-field emission patterns on the number of lasing
modes for stadium-shaped microlasers, revealing that the good
correspondence with the result of a ray simulation is obtained when
multiple modes are involved in lasing.

We carried out experiments for stadium-shaped microcavities with
strained InGaAsP multiple-quantum-well
separate-confinement-heterostructures, the detail of whose layer
structures has been reported in Ref. \cite{Fukushima05}.
The radius $R$ of the semi-circular parts of the stadium is
$15\,\mu$m.
The main difference between InGaAsP cavities and GaAs cavities used
in previous works \cite{Shinohara08, Fukushima04} is that the lasing
wavelength of the former, $\lambda\approx 1558$ nm, is almost twice
as large as that of the latter, $\lambda\approx 850$ nm.
That means, for fixed $R$, the dimensionless size parameter
$nkR=2n\pi R/\lambda$ for an InGaAsP cavity is almost half of that
for a GaAs cavity, since the effective refractive index of InGaAsP,
$n=3.23$, is almost same as that of GaAs, $n=3.3$.
For our InGaAsP cavity, $nkR$ becomes around $195$, for which we are
able to perform the wave calculation of cavity modes based on the
boundary element method \cite{Wiersig03} and found that the average
mode spacing is large enough to resolve individual modes with the
resolution of our spectrometer.

\section{Spectral characteristics}
\begin{figure}[t]
\centerline{\includegraphics[width=120mm]{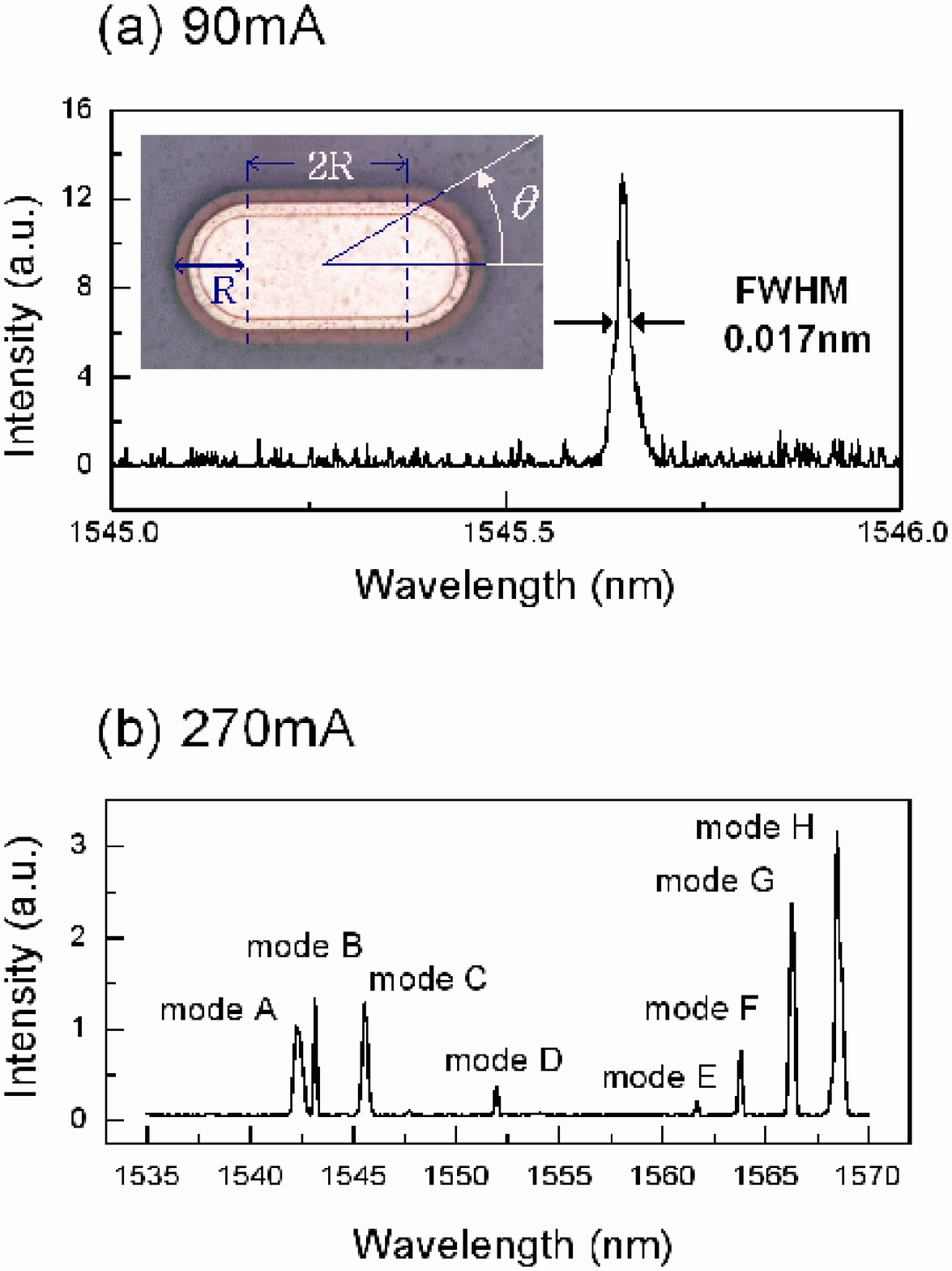}}
\caption{
Lasing spectra for the stadium-shaped InGaAsP microlaser. (a)
Single-mode lasing at pumping current $90$ mA. The inset shows the
microscope image of the Bunimovich stadium. (b) Multi-mode lasing at
$270$ mA.
} \label{fig:spectrum}
\end{figure}

We electrically pump the laser at room temperature using a pulsed
current of $30$ns-width with 1\% duty cycle.
In our experiment, setting the pulse width short is crucial for
maintaining thermal stability necessary for the achievement of
single-mode lasing.
The lasing threshold current is measured as $81$ mA.
The laser shows stable single-mode lasing for the pumping current
below 132 mA.
We show the spectrum data for single-mode lasing at the pumping
current $90$ mA in Fig. \ref{fig:spectrum}(a).
The achievement of the single-mode lasing is assured by two
experimental evidence.
One is that the FWHM of the peak in Fig. \ref{fig:spectrum}(a) is
$0.017$ nm, which is less than the wave numerical estimate of the
average mode spacing $0.055$ nm.
This estimate is based on the systematic analysis of cavity modes
found for the frequency range 60.94 $\leq \textrm{Re}\,kR \leq$
61.02 (i.e., 1545 nm $\leq \lambda \leq $ 1547 nm), where 34 modes
are detected.
When solving the equations, we assume transverse-electric(TE)
polarization, taking into account that the measured light is
TE-polarized.
The cavity modes are solutions of the Maxwell equations for a
stadium-shaped cavity without gain \cite{Wiersig03}.
The other evidence is that, below $132$ mA, the far-field emission
pattern does not change for the increase of the pumping current,
except for overall linear intensity growth.
At the pumping current $132$ mA, another lasing mode emerges and the
number of lasing modes grows continuously as the pumping current
increases.
For instance, at $270$ mA, we observed 8 lasing modes in the
spectrum as shown in Fig. \ref{fig:spectrum}(b).

\section{Far-field characteristics of single- and multi-mode lasing}
Figures 2(a) and 2(b) show the far-field patterns for single-mode
lasing at 100 mA and for multi-mode lasing at 270 mA, respectively.
We note that the intensity is normalized in both far-field patterns so
that the intensity integration over all angles becomes unity.
Moreover, spatial oscillations smaller than 0.8 degrees are smeared
out because of the resolution limit of our measurement.
In Figs. 2(a) and 2(b), we also plotted the result of a ray
simulation.
The ray simulation is based on the dynamics of an ensemble of rays
inside the cavity taking into account the ray emission at the cavity
boundary described by Fresnel's law.
The method of the ray simulation is explained in detail in
Refs. \cite{Shinohara06, Shinohara08}.
Comparing Figs. 2(a) and 2(b), we find that the oscillation amplitude
of the far-field pattern is larger in the single-mode lasing case than
in the multi-mode lasing case.
Moreover, in the multi-mode lasing case, the variation trend of the
far-field pattern can be well explained by the ray simulation result.

For better understanding of these experimental results, we performed
wave numerical calculation of far-field patterns.
We investigated 74 cavity modes with high quality factors sampled from
the frequency range, $60.00\leq \mbox{Re}\,kR\leq 61.40$ (i.e., 1535
nm $\leq \lambda \leq$ 1571 nm) and $-0.04\leq \mbox{Im}\,kR\leq 0.0$.
We emphasize that this frequency range corresponds to the range
relevant for our experiments.
Figure 2(c) is the far-field pattern of a high quality factor mode
whose frequency, $\mbox{Re}\,kR=60.96536$ (i.e., $\lambda=1545.9$ nm)
and $\mbox{Im}\,kR=-0.02385$, is close to the single-mode lasing
frequency, $\lambda=1545.7$ nm.
We see that the large-amplitude oscillation observed in the
single-mode lasing experiment is well reproduced in Fig. 2(c),
although we cannot make an exact correspondence of peak positions
between Fig. 2(a) and 2(c).
We note that the large-amplitude oscillation of a far-field pattern is
not a peculiar feature of the cavity mode shown in Fig. 2(c) but a
common feature of all the cavity modes.
The far-field pattern of the multi-mode lasing can be approximated by
the average of the far-field patterns of multiple cavity modes.
The pattern obtained by averaging 8 cavity modes is shown in
Fig. 2(d), where one can observe that, as in the multi-mode lasing
experiment, the oscillation amplitude becomes smaller and the variation
trend exhibits better agreement with the ray simulation result
compared to the single-mode case.

The correspondence of wave numerical results with the ray simulation
result can be confirmed more clearly for a larger cavity.
In Fig. 2(e), we plot wave numerical results for a stadium-shaped
cavity with $R=30\,\mu$m, which is twice as large as the one in our
experiments.
We plot in Fig. 2(e) the averaged far-field patten for 8 cavity modes
(green curve) and that for 54 cavity modes (black curve) together with
the ray simulation result (red broken curve), where one can clearly
observe the convergence of the wave numerical results onto the ray
simulation result by increasing the number of averaged modes.

\section{Spectral decomposition of a multi-mode lasing state}
\begin{figure}[t]
\centerline{\includegraphics[width=90mm]{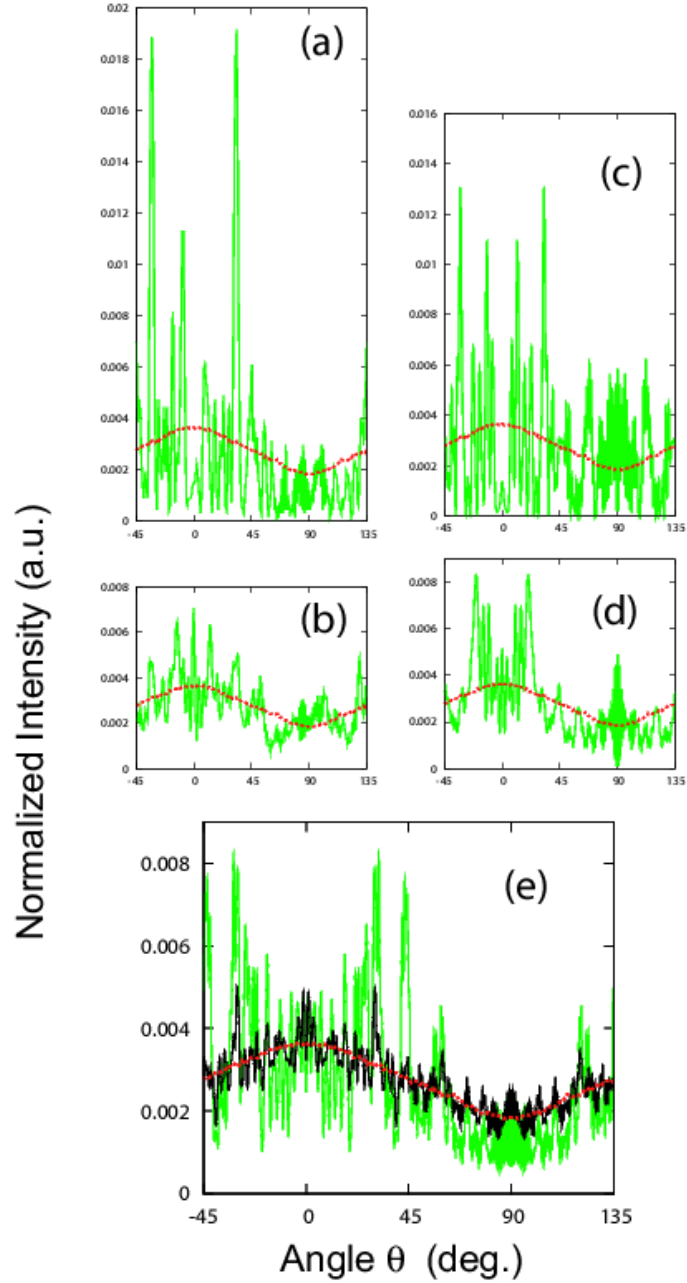}}
\caption{
Far-field patterns for experiments (green curves in (a) and (b)) and
for wave numerical simulations (green and black curves in (c), (d) and
(e)).
In (a)-(e), ray simulation data are plotted by a red broken curve.
(a) Experimental single-mode lasing data for the pumping current $100$ mA.
(b) Experimental multi-mode lasing data for $270$ mA.
(c) Wave calculation data of a single cavity mode.
(d) Wave calculation data of the average of 8 cavity modes.
These wave calculations are performed for a cavity with $R=15\,\mu$m,
which is the cavity size of our experiments.
(e) Wave calculation data for a cavity with $R=30\,\mu$m, where the
averaged far-field pattern for 8 cavity modes is plotted by a green
curve, while that for 54 cavity modes is plotted by a black curve.
} \label{fig:ffp}
\end{figure}

\begin{figure}[b]
\centerline{\includegraphics[width=120mm]{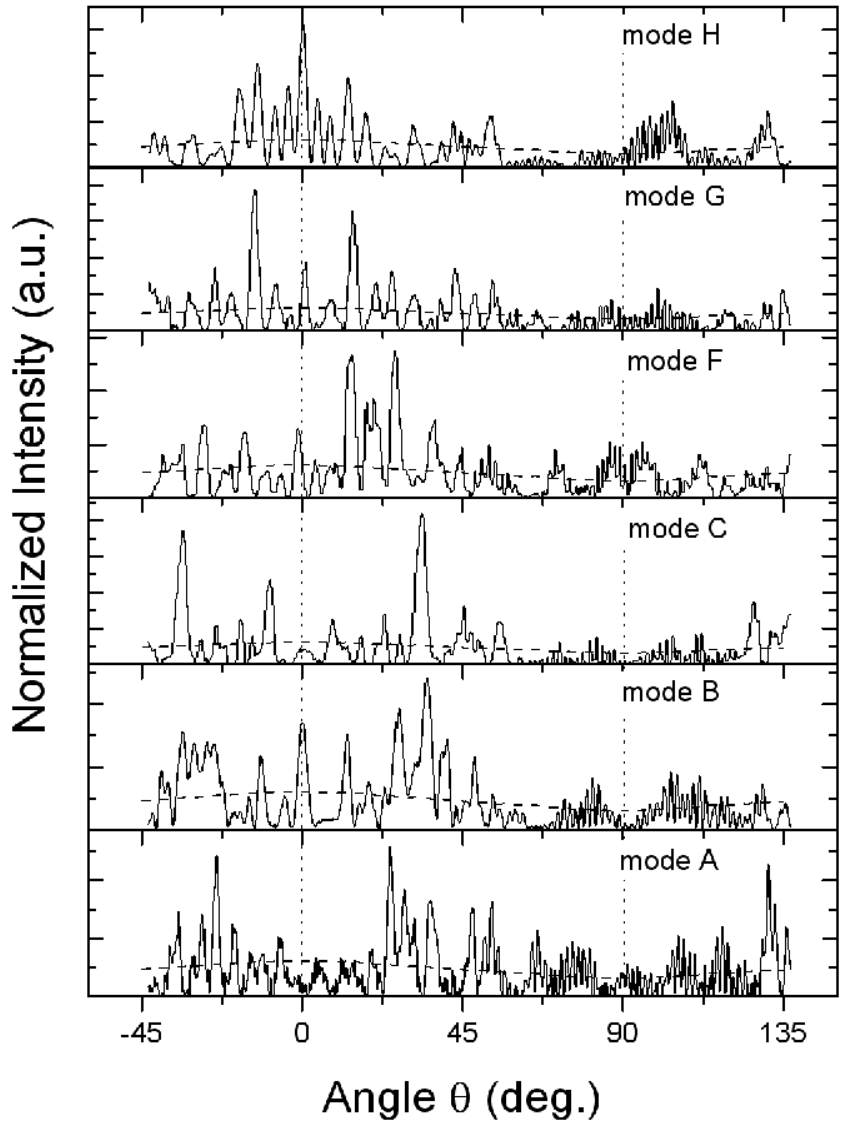}}
\caption{
Far-field patterns for the 6 dominant lasing modes for the pumping
current $270$ mA. The data are measured by a monochromator. The
labeling of the lasing modes corresponds to that in Fig.
\ref{fig:spectrum}(b). The result of a ray simulation is plotted in
dashed curves.
} \label{fig:ffp2}
\end{figure}

The experimental far-field patterns shown in Fig. \ref{fig:ffp}(a)
and \ref{fig:ffp}(b) are measured by a photodetector.
Using a monochromator instead, we can measure the far-field patterns
for individual lasing modes for the multi-mode lasing state for the
pumping current $270$ mA.
In Fig. \ref{fig:ffp2}, we plot the far-field patterns corresponding
to $6$ dominant lasing modes in Fig. \ref{fig:spectrum}(b), measured
by a monochromator(Ando AQ6317C) whose transmission band width is
$0.01$ nm.
We note that rapid spatial oscillations smaller than $0.24$ degrees
are smeared out in the measurement.
As expected from the analysis of the cavity modes in the previous
section, each of the lasing modes is found to have a less similar
pattern with the result of a ray simulation.
We note that mode C in Fig. \ref{fig:ffp2} is nothing but the mode
shown in Fig. \ref{fig:ffp}(a).
The agreement of the far-field pattern of mode C and that of the
single-mode lasing at $100$ mA convinces us the consistency between
the measurements by the photodetector and by the monochromator.
The existence of a variety of patterns in lasing modes as shown in
Fig. \ref{fig:ffp2} is contrasted with the previous work on the
quadrupole-deformed cavity \cite{Shim07}, where all lasing modes are
found to have far-field patterns closely corresponding to a ray
simulation result.
On the basis of our present result and previous theoretical analysis
\cite{Shinohara08}, we conclude that in general (i.e., for a generic
cavity shape with an arbitrary refractive index) the good
correspondence with a ray simulation result can not always be observed
for the far-field pattern of a single lasing mode, but can be observed
for the averaged far-field pattern such as those realized in
multi-mode lasing.

\section{Discussion}
In the experimental and wave numerical far-field patterns, one finds
that the spatial oscillations around the 90 degrees are more rapid
than those in the other range. This can be explained in the following
manner: From numerical analysis of cavity modes together with the
analysis of a ray model, we find that light is mostly emitted from the
two semi-circular parts of the stadium.  Thus, one can suppose that
far-field emission pattern around 90 degrees is the interference
pattern of the light coming from the two semi-circular parts. Since
the two light sources (i.e., two semi-circular parts) are located very
close, they result in rapid oscillation in the far-field emission
pattern. Approximating the distance of the two light sources as
$3R=45\,\mu \textrm{m}$, one can estimate the period of the
interference pattern as 2 degrees, which is consistent with both the
experimental data in Fig. 3 and the numerical data in Fig. 2(c). On
the other hand, the far-field pattern around, say, 0 degree is
supposed to be composed of light coming from one light source (i.e.,
one semi-circular part). Thus, for this angle range, one finds no
rapid oscillation due to the interference.

Lastly, we remark on the asymmetry of the far-field patterns observed
in Figs. \ref{fig:ffp}(a) and \ref{fig:ffp2}.
In theory, the far-field patterns of individual cavity modes obey
the C$_{2v}$ symmetry of the stadium shape.
However, some of the measured patterns in Figs. \ref{fig:ffp}(a)
and \ref{fig:ffp2} (e.g. mode F) exhibit noticeable asymmetries.
It is difficult to attribute these asymmetries to extrinsic
imperfections involved in the fabrication and measurement, since we
have also observed a rather symmetric pattern (e.g. mode A) for the
same cavity.
A plausible explanation is that an asymmetric lasing mode
corresponds not to a single cavity mode but multiple cavity modes
with their frequencies locked to a single frequency.
It has been demonstrated experimentally and numerically that
multiple frequency-locked modes can generate asymmetric emission
patterns \cite{SB}.

\section{Conclusion}
We studied spectral and far-field characteristics of stadium-shaped
InGaAsP microlasers.
We experimentally achieved single-mode lasing and showed that the
observed large-amplitude oscillations in the far-field pattern can be
reproduced by wave numerical calculation of cavity modes.
For the far-field pattern for multi-mode lasing, we found the
oscillation amplitude is suppressed and the global variation trend can
be well explained by a ray simulation result, which is also reproduced
by the wave numerical calculation.
These results are evidence for a hypothesis \cite{Shinohara08} that
the correspondence with a ray simulation becomes better as a result of
multi-mode lasing.

\section*{Acknowledgments}
The authors greatly thank Prof. Takehiro Fukushima for helpful
suggestions. The work at ATR was supported in part by the National
Institute of Information and Communications Technology of Japan.


%
%

%
%

%
%

\end{document}